\documentclass[a4paper,11pt]{article}
\usepackage{amsmath,amssymb,graphicx,bezier,citesort}
\newcommand{\nc}{\newcommand}
\nc{\rnc}{\renewcommand}
\nc{\be}{\begin{equation}}
\nc{\ee}{\end{equation}}
\nc{\bea}{\begin{eqnarray}}
\nc{\eea}{\end{eqnarray}}
\nc{\nn}{\nonumber}
\nc{\ch}{\cosh}
\nc{\sh}{\sinh}
\nc{\Z}{\overline{Z}}
\def\i{{\rm i}}
\def\e{{\rm e}}
\def\rhs{r.h.s.\,\,}
\def\refeq#1{(\ref{#1})}
\rnc{\th}{\tanh}
\rnc{\Im}{{\rm{Im}\,}}
\rnc{\Re}{{\rm{Re}\,}}
\nc{\mcK}{{\mathcal{K}}}
\nc{\mcL}{{\mathcal{L}}}
\nc{\ab}{\overline{a}}
\nc{\Ab}{\overline{A}}
\nc{\mfa}{{\mathfrak{a}}}
\nc{\mfab}{\overline{\mfa}}
\nc{\mfA}{{\mathfrak{A}}}
\nc{\mfAb}{\overline{\mfA}}
\nc{\mfsl}{{\mathfrak{sl}}}
\nc{\db}{\displaybreak[0]\\}
\nc{\sotimes}{\mathop{\otimes}_{s}}
\DeclareMathOperator*{\cp}{\stackrel{\curvearrowleft}{\prod}}

\numberwithin{equation}{section}

\begin{document}
%
\title{The thermal conductivity of the spin-1/2 $XXZ$ chain at arbitrary
temperature}
\author{
Andreas Kl\"umper\thanks{E-mail address:
kluemper@printfix.physik.uni-dortmund.de}  and 
Kazumitsu Sakai\thanks{E-mail address:
sakai@printfix.physik.uni-dortmund.de}\\\\
\it Theoretische Physik I, Universit\"at Dortmund, \\ 
\it Otto-Hahn-Str. 4, D-44221 Dortmund, Germany} 
\date{December 24, 2001} 
\maketitle
%
%
\begin{abstract}
Motivated by recent investigations of transport properties of strongly
correlated 1d models and thermal conductivity measurements of quasi 1d
magnetic systems we present results for the integrable spin-1/2 $XXZ$
chain.  The thermal conductivity $\kappa(\omega)$ of this model has
$\Re\kappa(\omega)=\tilde\kappa \delta(\omega)$, i.e. it is infinite for
zero frequency $\omega$. The weight $\tilde\kappa$ of the delta peak
is calculated exactly by a lattice path integral
formulation. Numerical results for wide ranges of temperature and
anisotropy are presented. The low and high temperature limits are
studied analytically.
\end{abstract}
%
%
\section{Introduction}
Transport properties of strongly correlated quantum systems have
recently attracted strong theoretical and experimental interest. For
the $XXZ$ chain as a quantum spin model (or alternatively as a lattice gas
model) the spin transport (electrical transport) was investigated
analytically \cite{Zotos98} by use of a method suggested in
\cite{FujKa98} and numerically \cite{NarMA98,AlGros01} with so far
inconclusive results about the finite temperature Drude weight. In
\cite{Kudo,Sologubenko00,Hess01}  the thermal conductivity of the 
quasi one-dimensional
magnetic system (Sr,Ca)$_{14}$Cu$_{24}$O$_{41}$ was investigated 
showing anomalous transport
properties along the chain directions. In \cite{Hess01} it was argued that
the large thermal conductivity can not be explained in terms of
phonons.

Here we want to use an approach as microscopic as possible and apply
Kubo's theory to the strongly correlated spin-1/2 $XXZ$ chain.  
This approach closely follows \cite{znp}. We explicitly
avoid the notion of particles and Boltzmann equations that take account
of scattering as is commonly done in Fermi liquid theory.  In strongly
correlated low-dimensional quantum systems there is neither a particle 
picture with finite quasi-particle weight nor do we restrict ourselves 
to low temperatures in comparison to some reference (``Fermi'') energy.
Rather on the opposite, we are interested in temperatures comparable to
the exchange energy of neighboring spins.

The Kubo formulas \cite{kubo,Mahan} 
are obtained within linear response theory and 
yield the (thermal) conductivity $\kappa$ relating the (thermal) current
$\mathcal{J}_{\rm E}$ to the (temperature) gradient $\nabla T$
\be
\mathcal{J}_{\rm E}=\kappa \nabla T, 
\end{equation}
where
\be
\kappa(\omega)=\beta\int_0^\infty dt \e^{-\i\omega t}\int_0^\beta
d\tau \langle \mathcal{J}_{\rm E}(-t-\i\tau) \mathcal{J}_{\rm
E}\rangle, 
\end{equation} 
and $\beta$ is the reciprocal temperature
$1/(k_{\rm B}T)$\footnote{In this paper we set $k_{\rm B}=1$.}. 
The current-current correlation function is to be
evaluated in thermal equilibrium and poses the main problem of our
work. However, as already pointed out in \cite{znp,lu,gm,Tsv,Frahm,Racz} the 
total thermal current operator $\mathcal{J}_{\rm E}$ commutes with the
Hamiltonian $\mathcal{H}$ of the $XXZ$ chain. Hence we find
\be
\kappa(\omega)=\frac 1{\i(\omega-\i\epsilon)}\beta^2
\langle\mathcal{J}_{\rm E}^2\rangle, \qquad (\epsilon\to0+),
\end{equation} 
with $\Re\kappa(\omega)=\tilde\kappa\delta(\omega)$ 
where
\be
\tilde\kappa=\pi\beta^2\langle\mathcal{J}_{\rm E}^2\rangle.
\end{equation}
As a consequence, the thermal conductivity
at zero frequency is infinite! The property of a non-decaying thermal
current is a feature of 1d systems in particular of integrable systems
with many integrals of motion (for a discussion see
\cite{z01}). We expect that additional interactions,
especially residual interchain couplings in quasi one-dimensional
materials will lead to a broadening of the delta peak, however with 
same weight. Instead of speculations about the possible scenarios 
of such a broadening we want to present results for the weight 
$\tilde\kappa$ as a function of temperature and anisotropy.

The paper is organized in the following way. In Sect.2 we discuss the
integrability of the $XXZ$ chain and identify the thermal current as
one of the integrals of motion. In Sect.3 we present our
method of calculation of thermodynamical quantities and discuss
the results. In Sect.4 we treat analytically the low and high temperature
limits and give a summary in Sect.5.

%
\section{Thermal current and conserved quantities}
Here we begin the microscopic approach to the heat conductivity of the
spin-1/2 quantum chain with a construction of the conserved
currents.  The
Hamiltonian of the $XXZ$ model on a periodic lattice of size $L$ is
\begin{align}
&\mathcal{H}=\sum_{k=1}^{L} h_{k k+1}, \nn \\
&h_{k k+1}=J\left(\sigma^{+}_{k}\sigma^{-}_{k+1}+
                        \sigma^{+}_{k+1}\sigma^{-}_{k}+
                \frac{\Delta}{2} \sigma^{z}_{k}\sigma^{z}_{k+1}
               \right),
\label{hamiltonian}
\end{align}
where $\sigma_k^{\pm}=(\sigma_k^{x}\pm i\sigma_k^{y})/2$ and 
$\sigma_k^{x}$, $\sigma_k^{y}$, $\sigma_k^{z}$
denote the usual Pauli matrices acting on the $k$th space.

In this paper we restrict ourselves to the critical 
regime $-1<\Delta\le 1$ where the system displays algebraically
decaying correlation functions at zero temperature.
The anisotropy parameter $\Delta$ is conveniently parameterized by
\be
\Delta=\cos\gamma,\qquad 0\le\gamma<\pi.
\end{equation}  
We will give our analytical results for the entire regime $0\le\gamma<\pi$
($-1<\Delta\le 1$), numerical results will be given for the repulsive range 
$0\le\gamma\le\pi/2$ ($0\le\Delta\le 1$).

Our first goal is the determination of the thermal current operator
$j^{\rm E}$. To this end we impose the continuity equation relating the time
derivative of the local Hamiltonian (interaction) and the divergence of
the current: $\dot h=-{\rm div}\, j^{\rm E}$.
The time evolution of the local Hamiltonian 
in~\eqref{hamiltonian} is obtained from the commutator with the total
Hamiltonian and the divergence of the local current on the lattice
is given by a difference expression
\be
\frac{\partial h_{k k+1}(t)}{\partial t}
=\i[\mathcal{H},h_{k k+1}(t)]=-\{j_{k+1}^{\rm E}(t)-j_{k}^{\rm E}(t)\}.
\end{equation}
Apparently the last equation is satisfied 
with a local thermal current operator $j^{\rm E}_{k}$
defined by
\be
j_k^{\rm E}=\i[h_{k-1 k},h_{k k+1}].
\end{equation}
In fact the total thermal current $\mathcal{J}_{\rm E}=
\sum_{k=1}^{L}j_{k}^{\rm E}$
\be
\mathcal{J}_{\rm E}=-\i J^2\sum_{k=1}^{L}
    \Bigl\{\sigma_k^{z}(\sigma_{k-1}^{+}\sigma_{k+1}^{-}-
           \sigma_{k+1}^{+}\sigma_{k-1}^{-})-
           \Delta(\sigma_{k-1}^{z}+\sigma_{k+2}^{z})
                 (\sigma_{k}^{+}\sigma_{k+1}^{-}-
                  \sigma_{k+1}^{+}\sigma_{k}^{-})\Bigr\},
\label{ec}
\end{equation}
commutes with the Hamiltonian, $[\mathcal{H},\mathcal{J}_{\rm E}]=0$,
as it is closely connected with a nontrivial conserved quantity 
derived from the underlying integrability of the model. 
To see this,  we introduce the transfer matrix
constructed from the $R$-matrix $R\in {\rm End}(V\otimes V)$
where $V$ denotes a two dimensional irreducible
module over the quantum algebra $U_q(\widehat{\mfsl}(2))$.
The nonzero 6 elements of the $R$-matrix with the spectral
parameter $v$ are given by
\be
R_{11}^{11}(v)=R_{22}^{22}(v)=\frac{[v+2]}{[2]},\quad
R_{12}^{12}(v)=R_{21}^{21}(v)=\frac{[v]}{[2]},\quad
R_{12}^{21}(v)=R_{21}^{12}(v)=1,
\end{equation}
where $[v]$ denotes $[v]=\sin(\gamma v/2)/\sin(\gamma/2)$
and the indices of the above relations can be interpreted as 
\begin{align}
&|1\rangle=|\uparrow\rangle,\qquad |2\rangle=|\downarrow\rangle,\nn \\
&R(v)|\alpha \rangle \otimes |\beta \rangle=
\sum_{\gamma,\delta}|\gamma \rangle \otimes |\delta \rangle
R_{\alpha \beta}^{\gamma \delta}(v).
\end{align}
The $R$-matrix satisfies the Yang-Baxter equation (YBE)
\be
R_{12}(u-v)R_{13}(u)R_{23}(v)=R_{23}(v)R_{13}(u)R_{12}(u-v).
\end{equation}
Due to the YBE the transfer matrix 
$T(v)=\cp_{k}R_{k k+1}(v)$
is commutative with respect to  different spectral
parameter $v$ and $v^{\prime}$, i.e 
\be
[T(v),T(v^{\prime})]=0.
\label{commute}
\end{equation}
As is well known, $\ln T(v)$ is the generating function for 
the conserved quantities
\be
\mathcal{J}^{(n)}=\left(\frac{\partial}{\partial v}\right)^n
\ln T(v) \Big|_{v=0}.
\label{generator}
\end{equation}
In particular the Hamiltonian~\eqref{hamiltonian} and the thermal
current~\eqref{ec} are expressed in terms of
 $\mathcal{J}^{(1)}$ and $\mathcal{J}^{(2)}$, respectively.
Explicitly they read
\begin{align}
&\mathcal{H}=\frac{2J\sin\gamma}{\gamma}\mathcal{J}^{(1)}
             -\frac{J L}{2} \Delta,\nn \\
&\mathcal{J}_{\rm E}=
             \i \left(\frac{2J\sin\gamma}{\gamma}\right)^2
              \mathcal{J}^{(2)}+
              \i {J^2 L}.
\label{conserve} 
\end{align}
Due to the commutativity~\eqref{commute}, every 
operator $\mathcal{J}^{(n)}$ commutes with the Hamiltonian
$\mathcal{H}$.
%
\section{Thermal conductivity}
%
Our goal is the calculation of the second moment of the thermal
current. Quite generally the expectation values of conserved
quantities may be calculated by use of a suitable generating
function. As such we want to define a modified partition function
\be
Z={\rm Tr}\,\exp(-\beta\mathcal{H}+\lambda \mathcal{J}_{\rm E}),
\label{part}
\end{equation}
from which we find the expectation values by derivatives with respect to 
$\lambda$ at $\lambda=0$
\be
\frac{\partial}{\partial \lambda}\ln Z \Big|_{\lambda=0}
=\langle \mathcal{J}_{\rm E} \rangle=0,\qquad
\left(\frac{\partial}{\partial \lambda}\right)^2\ln Z \Big|_{\lambda=0}
=\langle \mathcal{J}_{\rm E}^2 \rangle
-\langle \mathcal{J}_{\rm E} \rangle^2
=\langle \mathcal{J}_{\rm E}^2 \rangle,
\end{equation}
where we have used that the expectation value of the thermal current
in thermodynamical equilibrium is zero. 

Instead of $Z$ we will find it slightly more convenient to work with 
a partition function
\be
\Z={\rm Tr}\,\exp(-\lambda_1\mathcal{J}^{(1)}-\lambda_n\mathcal{J}^{(n)}).
\label{part2}
\end{equation}
With view to \refeq{conserve} we choose
\be
\lambda_1=\beta\frac{2J\sin\gamma}{\gamma},\qquad
\lambda_2=-\i \lambda \left(\frac{2J\sin\gamma}{\gamma}\right)^2,
\end{equation}
and obtain the desired expectation values from $\Z$
\be
\langle \mathcal{J}_{\rm E}^2 \rangle=
\left(\frac{\partial}{\partial \lambda}\right)^2\ln\Z \Big|_{\lambda=0}.
\end{equation}

We can deal with $\Z$ rather easily. Consider the trace of a product 
of $N$ row-to-row transfer matrices $T(u_j)$ with some spectral parameters
$u_j$ close to zero, but still to be specified, and the $N$th power of
the inverse of $T(0)$
\begin{align}
Z_N&={\rm Tr}\,\left[T(u_1)\cdot\, \dotsb \,\cdot T(u_N)\cdot T(0)^{-N}\right]\cr
&={\rm Tr}\,\exp\left(\sum_j[\ln T(u_j)-\ln T(0)]\right).
\end{align}
Now it is a standard exercise in arithmetic to devise a sequence of
$N$ numbers $u_1$,...,$u_N$ (actually $u_j=u_j^{(N)}$) such that 
\be
\lim_{N\to\infty}\sum_j[f(u_j)-f(0)]=
-\lambda_1\frac{\partial}{\partial v}f(v)\Big|_{v=0}
-\lambda_n\left(\frac{\partial}{\partial v}\right)^nf(v)\Big|_{v=0}.
\end{equation}
We only need the existence of such a sequence of numbers, the precise
values are of no importance. In the limit $N\to\infty$ we note
\be
\lim_{N\to\infty}Z_N=\Z.
\end{equation}

We can proceed along the established path of the quantum transfer
matrix (QTM) formalism 
\cite{MSuzPB,InSuz,InSuz2,Koma,KlumTH,TakQT,Mizu}
and derive the partition function $Z_N$ in the
thermodynamic limit $L\to\infty$
\be
\lim_{L\to\infty}Z_N^{1/L}=\Lambda,
\end{equation}
where $\Lambda$ is the largest eigenvalue of the QTM. The
integral expression for $\Lambda$ reads
\begin{align}
&\ln\Lambda=\sum_j[e(u_j)-e(0)]
+\int_{-\infty}^{\infty}K(v)\ln[\mfA(v)\mfAb(v)]dv,
&K(v)=\frac{1}{4\cosh{\frac{\pi}{2}v}},
\label{largest}
\end{align}
with some function $e(v)$ given in \cite{KlumTH}. In the limit $N\to\infty$
the first term on the \rhs of the last equation turns into
\be
\lim_{N\to\infty}\sum_j[e(u_j)-e(0)]=
-\lambda_1\frac{\partial}{\partial v}e(v)\Big|_{v=0}
-\lambda_n\left(\frac{\partial}{\partial v}\right)^ne(v)\Big|_{v=0},
\end{equation}
a rather irrelevant term as it is linear in $\lambda_1$ and $\lambda_n$,
and therefore the second derivatives with respect
to $\lambda_1$ and $\lambda_n$ vanish.
The functions $\mfA(v)$ and $\mfAb(v)$ are determined from
the following set of non-linear integral equations (NLIEs):
\begin{align}
&\ln\mfa(v)=\sum_j[\varepsilon_0(v-\i u_j)-\varepsilon_0(0)]+
            \kappa\ast\ln\mfA(v)-\kappa\ast\ln\mfAb(v+2i),\nn \db
&\ln\mfab(v)=\sum_j[\varepsilon_0(v-\i u_j)-\varepsilon_0(0)]+
            \kappa\ast\ln\mfAb(v)-\kappa\ast\ln\mfA(v-2i),\nn \db
&\mfA(v)=1+\mfa(v),\qquad \mfAb(v)=1+\mfab(v).
\label{nlie}
\end{align}
with a function $\varepsilon_0(v)$ given in terms of hyperbolic
functions \cite{KlumTH}. The symbol $\ast$ denotes the convolution $f\ast g(v)=
\int_{-\infty}^{\infty} f(v-v^{\prime})g(v)dv$ and the function 
$\kappa(v)$ is defined by
\be
\kappa(v)=\frac{1}{2\pi}\int_{-\infty}^{\infty}
          \frac{\sh\left(\frac{\pi}{\gamma}-2\right)k}
                {2\ch k
                 \sh\left(\frac{\pi}{\gamma}-1\right)k} e^{ikv}dk.
\label{bulk}
\end{equation}
Again, the summations in \refeq{nlie} can be simplified in the
limit $N\to\infty$
\be
\lim_{N\to\infty}\sum_j[\varepsilon_0(v-\i u_j)-\varepsilon_0(v)]=
-\lambda_1\underbrace{
\left(-\i\frac{\partial}{\partial v}\right)\varepsilon_0(v)
}_{=:\varepsilon_1(v)}
-\lambda_n\underbrace{\left(-\i\frac{\partial}{\partial v}
\right)^n\varepsilon_0(v)}_{=:\varepsilon_n(v)}.
\label{drivter}
\end{equation}
where the first function can be found in \cite{KlumTH} and is simply
\be
\varepsilon_1(v)=2\pi K(v)=
\frac{\pi}{2\ch\frac{\pi}{2}v},
\end{equation}
and hence the second function is
\be
\varepsilon_n(v)=\left(-\i\frac{\partial}{\partial
v}\right)^{n-1}\varepsilon_1(v).
\end{equation}
We like to note that the structure of the driving term \refeq{drivter}
appearing in the NLIE \refeq{nlie} reflects the structure of the
generalized Hamiltonian in the exponent on the \rhs of \refeq{part2}. 
We could have given an alternative derivation
of the NLIE along the lines of the thermodynamic Bethe Ansatz (TBA). 
In such an approach the driving term is typically the one-particle energy 
corresponding to the generalized Hamiltonian. Hence it has contributions
due to the first as well as the $n$th logarithmic derivative of the
row-to-row transfer matrix, i.e. the terms $\varepsilon_1$ and $\varepsilon_n$.

\begin{figure}
\begin{center}
\includegraphics[width=0.85\textwidth]{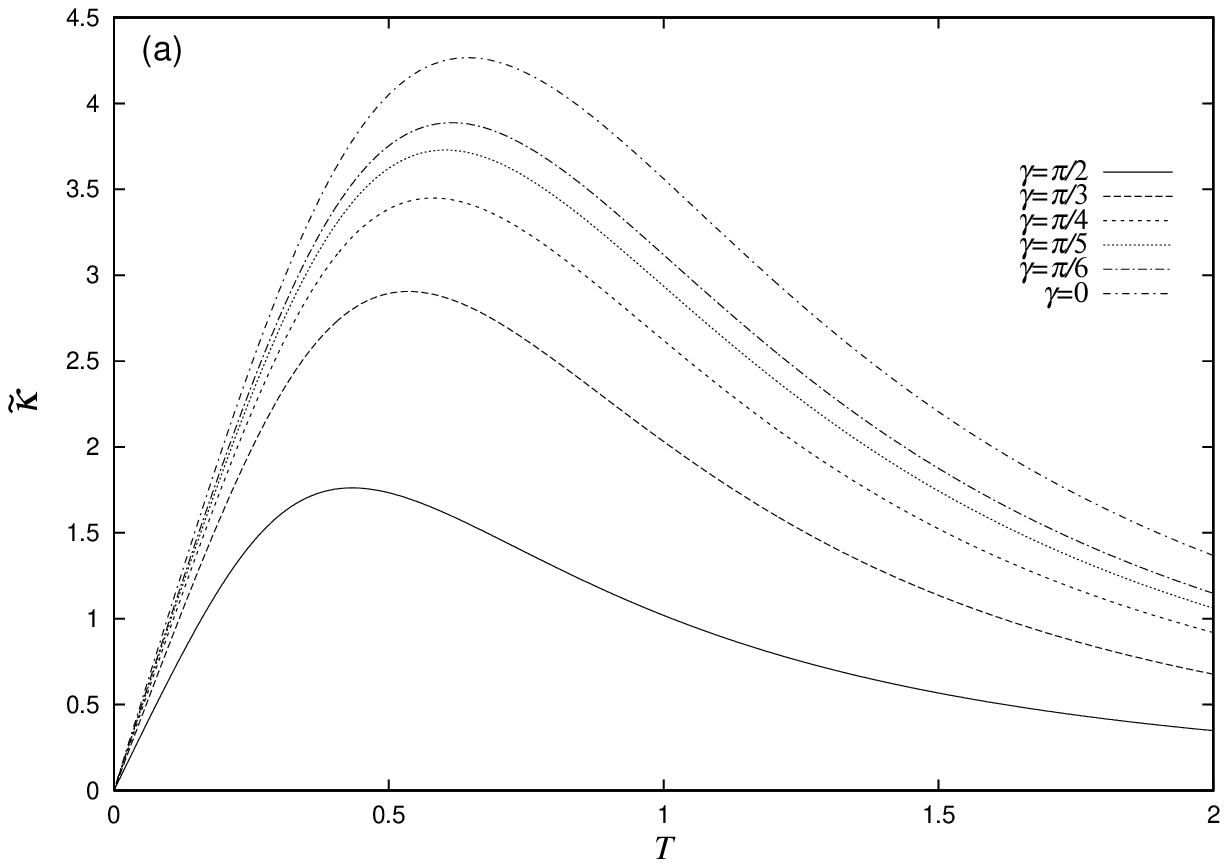}
\includegraphics[width=0.85\textwidth]{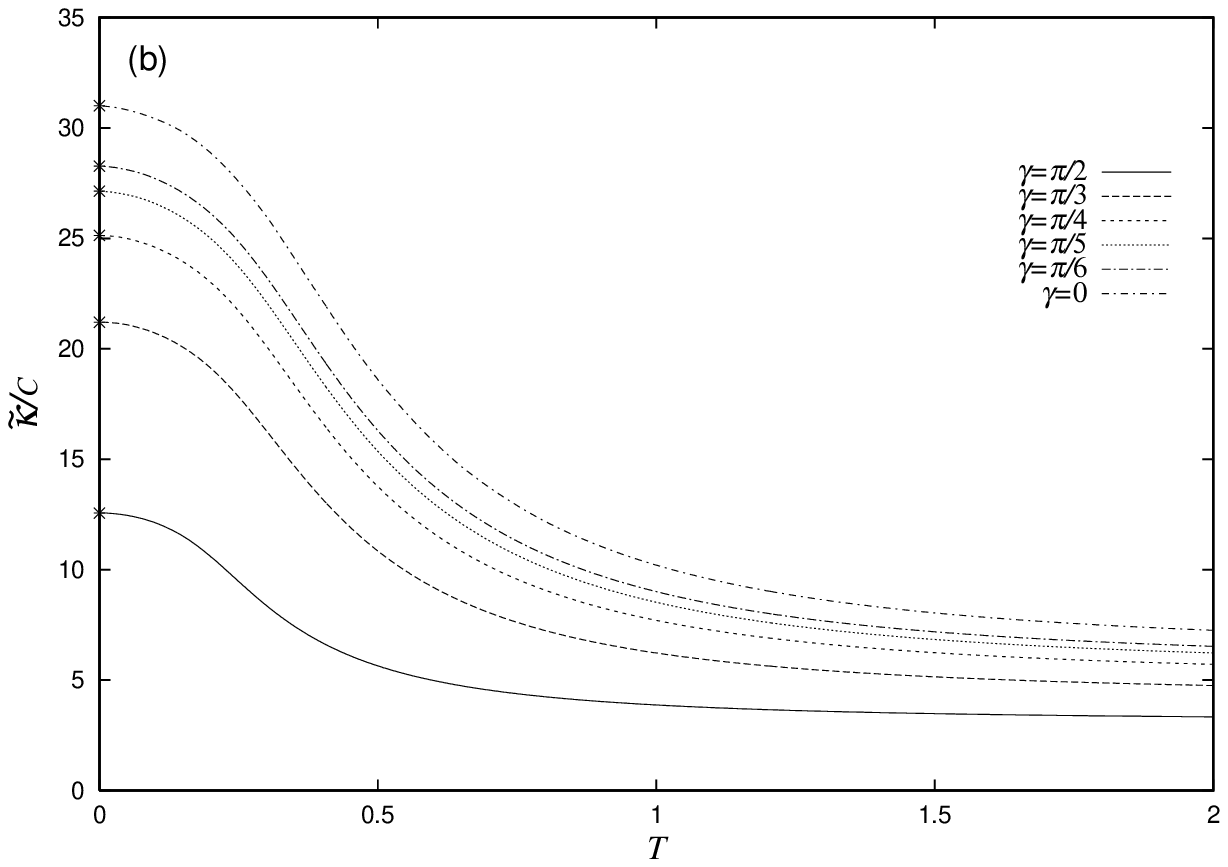}
\end{center}
\caption{(a) Illustration of numerical results for the thermal conductivity
$\tilde\kappa$ as a function of temperature $T$ for various anisotropy 
parameters $\gamma=0, \pi/6, \pi/5, \pi/4, \pi/3, \pi/2$.
(b) Depiction of the ratio of thermal conductivity and specific heat
$\tilde\kappa/c$ as a function of temperature. The analytic result \eqref{ck}
in the low temperature limit is also depicted by the symbol $*$.
}
\label{f1}
\end{figure}

In Fig.\ref{f1} (a) we show $\tilde\kappa(T)$ for various anisotropy
parameters $\gamma$. Note that $\tilde\kappa(T)$ has linear $T$ dependence
at low temperatures. 
%
In the free fermion case ($\gamma=\pi/2$) we observe 
that the ratio of the thermal conductivity $\kappa$ to the 
electrical conductivity $\sigma$ 
obeys the Wiedemann-Franz law (see the next section).
There is a finite temperature maximum at roughly 
half of the temperature of the maximum in the specific heat $c(T)$ data.
At high temperatures $\tilde\kappa(T)$ behaves like $1/T^2$. This and
the low temperature asymptotics will be studied analytically in the next
section.

In general the data of $\tilde\kappa(T)$ show a much stronger variation with
the anisotropy $\gamma$ than the specific heat $c(T)$ data. In Fig.\ref{f1} (b)
we show the ratio $\tilde\kappa(T)/c(T)$ which have finite low and high
temperature limits and strongly depend on the anisotropy parameter.
%
\section{Low and high temperature limits}
In certain limits we can analytically evaluate the NLIE. For high
temperatures the NLIE linearize and hence can be solved. At low temperatures
the NLIE remain non-linear, but the driving terms simplify. Then the
symmetry of the integration kernel allows for an evaluation of the
physically interesting quantities by avoiding the explicit solution of
the NLIE.

%
\subsection{Low temperature limit}
We consider the low-temperature behavior of the current-current
correlation function of the thermal current $\mathcal{J}_{\rm E}$ by
making use of the dilogarithm trick.
The functions $\mfa(v)$ and $\mfab(v)$ in the NLIEs~\eqref{nlie}
exhibit a crossover behavior
\begin{alignat}{3}
&\mfa(v)\ll 1, &\quad& \mfab(v)\ll 1  &\quad&
       \text{for \quad $-\mathcal{K}_{-}<v<\mathcal{K}_{+}$}, \nn \db
&\mfa(v)\simeq 1,&\quad& \mfab(v)\simeq 1 &\quad&
                      \text{for \quad $v<-\mathcal{K}_{-}$, 
                                      $\mathcal{K}_{+}<v$},
\end{alignat}
where
\be
\mathcal{K}_{\pm}=\frac{2}{\pi}\ln\left\{
                  \frac{2\pi J\sin\gamma}{\gamma}
                  \left(\beta\pm\frac{\lambda 
                  \pi J\sin\gamma}{\gamma}\right)\right\}.
\end{equation}
We introduce the scaling functions
\be
a_{\pm}(v)=\mfa\left(\pm \frac{2}{\pi} v\pm \mathcal{K}_{\pm}
\right), \quad
\ab_{\pm}(v)=\mfab\left(\pm \frac{2}{\pi} v\pm \mathcal{K}_{\pm}
\right),
\end{equation}
and similarly the functions
$A_{\pm}$ ($\Ab_{\pm}$) for $\mfA$ ($\mfAb$).
{}From the NLIEs~\eqref{nlie}, one sees 
the scaling functions satisfy the 
following ``scaled'' equations in the limit 
$\beta\to \infty$:
\begin{align}
&\ln a_{\pm}(v)=-e^{-v}+
         \kappa_1\ast\ln A_{\pm}(v)-\kappa_{2\pm}\ast\ln\Ab_{\pm}(v),\nn \db
&\ln\ab_{\pm}(v)=-e^{-v}+
            \kappa_1\ast\ln\Ab_{\pm}(v)-\kappa_{2\mp}\ast\ln A_{\pm}(v), \nn \db
&\kappa_1(v)=\frac{2}{\pi}\kappa\left(\frac{2v}{\pi}\right),
\quad
\kappa_{2\pm}(v)=\frac{2}{\pi}\kappa\left(\frac{2v}{\pi}\pm 2i\right).
\label{scaling}
\end{align}
In this limit, the integrals of the functions
$\mfA$, $\mfAb$ in eq.~\eqref{largest} can be
written as
\begin{align}
\ln\Lambda&
=\int_{-\infty}^{\infty}K(v)\ln[\mfA(v)\mfAb(v)]dv \nn \\
&=\frac{\gamma^2}
       {2\pi^2 J\sin\gamma(\beta \gamma+\lambda\pi J \sin\gamma)}
           \int_{-\infty}^{\infty}e^{-v}(\ln A_{+}(v)+
            \ln \Ab_{+}(v))dv  \nn \\
&+\frac{\gamma^2}
            {2\pi^2J\sin\gamma(\beta \gamma-\lambda\pi J \sin\gamma)}
         \int_{-\infty}^{\infty}e^{-v}(\ln A_{-}(v)+
            \ln \Ab_{-}(v))dv .
\label{asy1}
\end{align}
The right hand side of the above equation is evaluated as
follows.
(A) After taking the derivative of the first and second equation in
\eqref{scaling}, we multiply them by $\ln A_{\pm}(v)$ and 
$\ln\Ab_{\pm}(v)$, respectively and take the summation of
them with respect to  each side.
(B) Next we multiply~\eqref{scaling} by $[\ln A_{\pm}(v)]^{\prime}$
and $[\ln \Ab_{\pm}(v)]^{\prime}$, respectively and take the 
summation.
Next we subtract the resultant equation of (B) from the one of
(A). After integrating over $v$, we obtain
\be
D_{\pm}=2\int_{-\infty}^{\infty}e^{-v}[\ln A_{\pm}(v)+\ln \Ab_{\pm}(v)] dv,
\label{asy2}
\end{equation}
where
\begin{align}
D_{\pm}&=\int_{-\infty}^{\infty}\Bigl(
        \ln A_{\pm}(v) \frac{d}{dv}\ln a_{\pm}(v)+
        \ln\Ab_{\pm}(v)\frac{d}{dx}\ln \ab_{\pm}(v)\nn \\
       &\qquad-\ln a_{\pm}(v) \frac{d}{dv}\ln A_{\pm}(v)-
        \ln\ab_{\pm}(v)\frac{d}{dx}\ln \Ab_{\pm}(v) \Bigr) dv \nn \\
       &=\int_{a_{\pm}(-\infty)}^{a_{\pm}(\infty)}
        \left(\frac{\ln(1+a)}{a}-\frac{\ln a}{1+a}\right)da+
        \int_{\ab_{\pm}(-\infty)}^{\ab_{\pm}(\infty)}
        \left(\frac{\ln(1+\ab)}{\ab}-\frac{\ln \ab}{1+\ab}\right)d\ab.
\end{align}
The quantities $D_{\pm}$ can be expressed in terms of Roger's 
dilogarithm $\mathcal{L}(v)$
\begin{align}
D_{\pm}&=2\mcL\left(\frac{a_{\pm}(\infty)}{1+a_{\pm}(\infty)}\right)+
        2\mcL\left(\frac{\ab_{\pm}(\infty)}{1+\ab_{\pm}(\infty)}\right)\nn \\
     &\quad-2\mcL\left(\frac{a_{\pm}(-\infty)}{1+a_{\pm}(-\infty)}\right)-2
      \mcL\left(\frac{\ab_{\pm}(-\infty)}{1+\ab_{\pm}(-\infty)}\right),\nn\db
\mcL(v)&=-\frac{1}{2}\int_{0}^{v}\left(\frac{\ln(1-x)}{x}+
          \frac{\ln x}{1-x}\right) dx.
\end{align}
Using the asymptotic value of the scaling functions
$a_{\pm}(\pm\infty)=\ab_{\pm}(\pm \infty)=1$ and
substituting eq.~\eqref{asy2} for eq.~\eqref{asy1},
we arrive at
\be
\ln\Lambda=\frac{\gamma^2 }
            {12\beta J\sin\gamma}\left(\frac{1}
              {\beta\gamma+\lambda\pi J\sin\gamma}+\frac{1}
              {\beta\gamma-\lambda\pi J\sin\gamma}\right).
\end{equation}
Here we have used the identity $\mcL(v)+\mcL(1-v)=\pi^2/6$.
Hence the low-temperature asymptotics of the current-current
correlation function is evaluated to
\be
\langle\mathcal{J}_{{\rm E}}^2\rangle\simeq
\frac{J\pi^2\sin\gamma}{3\beta^3\gamma}+O\left(\frac{1}{\beta^4}\right).
\end{equation}
{}From this result we see that $\tilde\kappa(T)$ is linear in $T$ at low $T$
\be
\tilde{\kappa}(T)\simeq
\frac{\pi^2}{3}vT,
\end{equation}
with $v=J\pi\sin\gamma/\gamma$ the velocity of the elementary excitations.
{}From the low temperature behavior of the specific heat 
$c(T)=(\pi/3v)T$  we find
\be
\frac{\tilde\kappa(T)}{c(T)}\to\pi v^2.
\label{ck}
\end{equation}
%
Finally we want to compare the thermal conductivity $\kappa$ and the 
electrical conductivity $\sigma$ 
which in the low temperature limit can be described by 
the Drude weight $D_{\rm c}$ \cite{SS}
\be
\Re \sigma(\omega)=2\pi D_{\rm c} \delta(\omega),\qquad
D_{\rm c}=\frac{v}{4(\pi-\gamma)}.
\end{equation}
This yields
\be
\frac{\kappa}{\sigma} \simeq \frac{2}{3}\pi(\pi-\gamma) T, \qquad (T\to 0),
\end{equation}
which in the free fermion case ($\gamma=\pi/2$) gives the Wiedemann-Franz law.

%
%
\subsection{High temperature limit}
In the high temperature limit ($\beta\to0$),
the auxiliary functions satisfy the following 
integral equations linear in $(\partial/\partial\lambda)^2\ln\mfA$
\begin{align}
\left(\frac{\partial}{\partial \lambda}\right)^2\ln\mfA(v)
&=
\left(\frac{\partial}{\partial\lambda}\ln\mfA(v)\right)^2+
\frac{1}{2}\kappa\ast
\left(\frac{\partial}{\partial\lambda}\right)^2\ln\mfA(v)-
\frac{1}{2}\kappa\ast
\left(\frac{\partial}{\partial\lambda}\right)^2\ln\mfAb(v+2\i),\nn \\
\left(\frac{\partial}{\partial \lambda}\right)^2\ln\mfAb(v)
&=
\left(\frac{\partial}{\partial\lambda}\ln\mfAb(v)\right)^2+
\frac{1}{2}\kappa\ast
\left(\frac{\partial}{\partial\lambda}\right)^2\ln\mfAb(v)-
\frac{1}{2}\kappa\ast
\left(\frac{\partial}{\partial\lambda}\right)^2\ln\mfA(v-2\i).
\end{align}
Here we have used the high temperature asymptotics 
$\mfa(v), \mfab(v)\simeq 1$.
By use of the dressed function formalism, we obtain the identity
\begin{align}
\left(\frac{\partial}{\partial \lambda}\right)^2\ln\Lambda
\biggr|_{\lambda=0}&=\frac{1}{2\pi}
\int_{-\infty}^{\infty}\varepsilon_1(v)
\left(\frac{\partial}{\partial \lambda}\right)^2\ln\mfA(v)\mfAb(v)
\biggr|_{\lambda=0}dv \nn \\
&=-\frac{\gamma}{8\pi J \sin\gamma}\int_{-\infty}^{\infty}
\frac{\partial\mfa(v)}{\partial\beta}\left(
\frac{\partial\mfa(v)}{\partial\lambda}\right)^2\biggr|_{\lambda=0}dv.
\end{align}
The integrand in the above equation is
found analytically
\begin{align}
\frac{\partial\mfa(v)}{\partial \beta}\biggr|_{\lambda=0}&=
\frac{J\sin^2\gamma}{2\sinh\frac{\gamma}{2}(v+\i)}
\left(\frac{1}{\sinh\frac{\gamma}{2}(v+3\i)}-
      \frac{1}{\sinh\frac{\gamma}{2}(v-\i)}\right), \nn \\
\frac{\partial\mfa(v)}{\partial \lambda}\biggr|_{\lambda=0}&=
-\frac{\partial}{\partial v}
\left(\frac{\partial\mfa(v)}{\partial \beta}\right)\biggr|_{\lambda=0}.
\end{align}
By use of these explicit expressions we obtain the high temperature limit of 
$\tilde{\kappa}(T)$
\be
\tilde{\kappa}(T)\simeq
\frac{J^4}{4}\left(3+\frac{\sin3\gamma}{\sin\gamma}\right) \beta^2+
O(\beta^3).
\end{equation}
This result generalizes the known results for
the special cases $\gamma=\pi/2$ (free fermion model) in \cite{Niem}
and $\gamma=0$ (isotropic Heisenberg chain) in \cite{nz}.
%

\section{Summary and discussion}
%
In this work we have presented a method for the calculation of
the thermal conductivity $\kappa(\omega)$ of the integrable spin-1/2
$XXZ$ chain. The investigation of this system is drastically 
simplified in comparison to other systems as here the thermal current 
is a conserved quantity. This led to $\Re\kappa(\omega)=
\tilde\kappa \delta(\omega)$ where the weight $\tilde\kappa$ was
calculated for arbitrary temperature and various anisotropy parameters
in the repulsive, critical regime of the $XXZ$ model. 

We like to comment on possible generalizations of our
investigation.  First of all, the computation of $\kappa(q,\omega)$,
i.e. the thermal conductivity for a thermal current operator
$\mathcal{J}_{\rm E}(q)$ with non-zero momentum $q$, would be
desirable. Unfortunately, it is only the case $q=0$ that allows for an
analytical approach. On the other hand, for small values of $q$ a
behavior of $\kappa(q,\omega)$ similar to the $q=0$ case is
expected with the $\delta(\omega)$ factor to be replaced by
$\delta(\omega-vq)$ where $v$ is the velocity of the elementary
excitations. Finally, the investigation of the thermal conductivity
of the general $XXZ$ chain with arbitrary anisotropies in
the easy plane (gapless) and easy axis (gapped) regimes should be
possible. This will be reported elsewhere.

%
\section*{Acknowledgments}
We would like to thank B. B\"uchner, C. Gros, D. Rainer, and X. Zotos 
for stimulating discussions.
The authors acknowledge financial support by the Deutsche 
Forschungsgemeinschaft under grants No.~Kl 645/3-3, 645/4-1 and the
Schwerpunktprogramm SP1073.
%

%
%
%

%
\end{document}